\documentclass[conference]{IEEEtran}

\usepackage{color}
\usepackage{graphicx}
\usepackage{amsmath}
\usepackage{multicol}
\usepackage{hyperref}
\usepackage{xcolor}

\title{ProxiTrak: Intelligent Enablement of Social Distancing \& Contact Tracing for a Safer Workplace in the New Normal}

\author{Authors: Vivek Chandel, Snehasis Banerjee | Mentor: Avik Ghose | Affiliation: TCS Research}
\begin{document}
\maketitle

\section{Brief Project Synopsis}
\label{section_brief}

This paper describes an innovative solution that enables the enterprises to bring their associates (or employees) back to physical workspaces for critical operations in a safe manner in the wake of current COVID-19 pandemic.

\begin{itemize}
\item \textbf{Using an on-device app,} our solution guides associates towards a real-time social distancing by leveraging ubiquity of BLE-enabled personal mobile devices/smart wearables for detecting whether any two associates have come within a minimum prescribed social distance using a device agnostic Machine Learning model. On-device real-time intervention notifications are generated for proximity events, thereby limiting the contagion to a great extent.
\item \textbf{On the server side,} we implement feature-rich graph analytics providing real-time insights into the risk profiling of associates, multi-level social connectivity, contagion risk propagation and identification of contagion clusters.
\item Our solution also supports use of employee access management infrastructure, and an external BLE infrastructure (if available) in regions within a workspace where personal devices may not be allowed.
\end{itemize}

Backed by multiple Tier-I publications \cite{chandel2020}\cite{banerjee2020}\cite{chandel2016}\cite{ghose2013}, patents \cite{chandel2020patent}\cite{banerjee2020patent}\cite{ghose2015patent}\cite{ghose2016patent}, successful pilots in real factory environments and workspaces, and integration with our in-house Workspace Resilience product (released online for use by over 4 lac associates), our technology is ready to be implemented at a larger scale by any industry. In short and medium term, the use of this innovation is employee as well as workplace safety. In long term, the solution can be integrated with social infrastucture of the enterprise and for studying general social interactions and generating social cluster anaytics in workspaces.

\begin{figure}
    \centering
    \includegraphics[width=0.8\columnwidth]{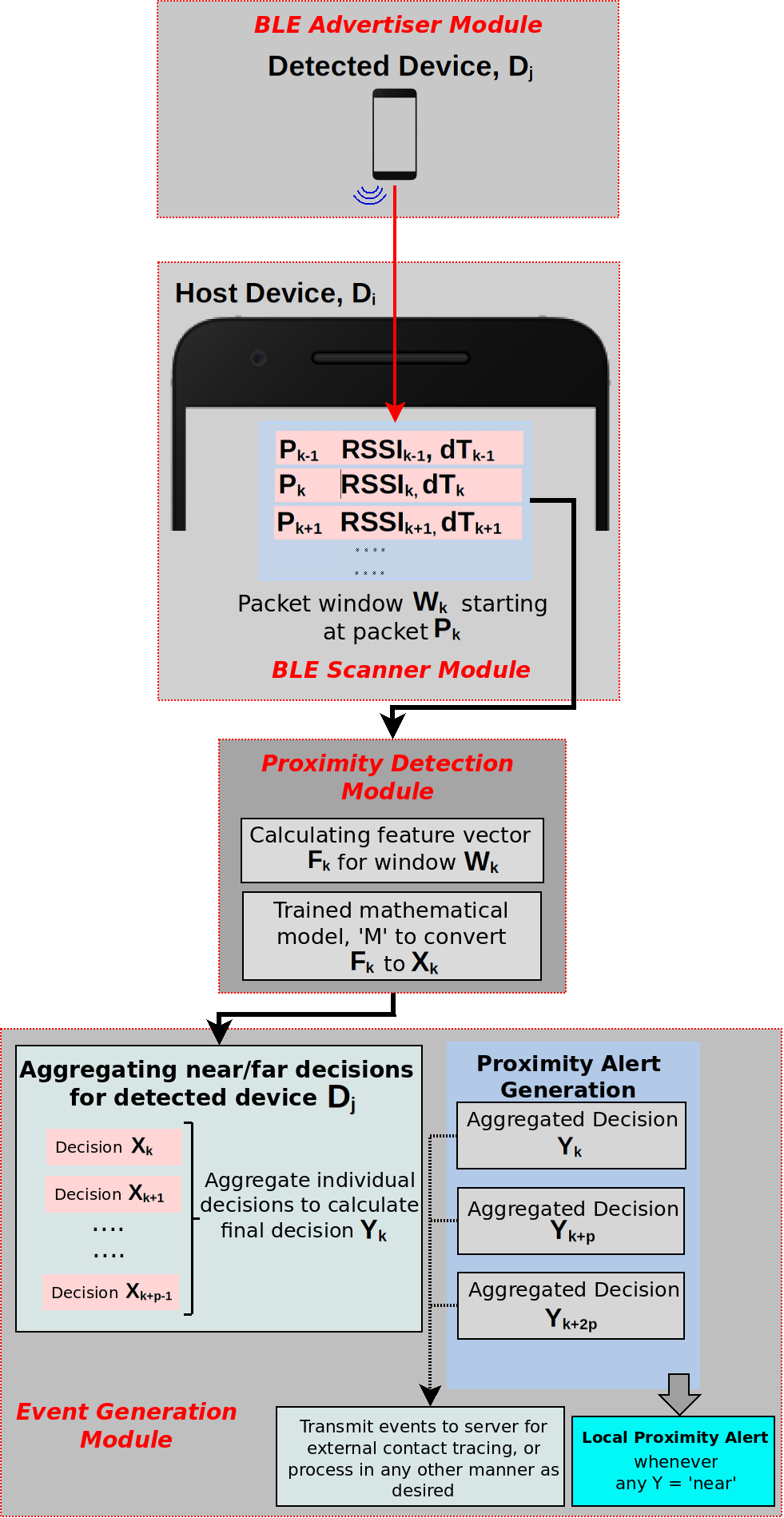}
    \caption{ProxiTrak Mobile App Architecture.}
    \label{fig_app_arch}
\end{figure}

\begin{figure}
    \centering
    \includegraphics[width=0.85\columnwidth]{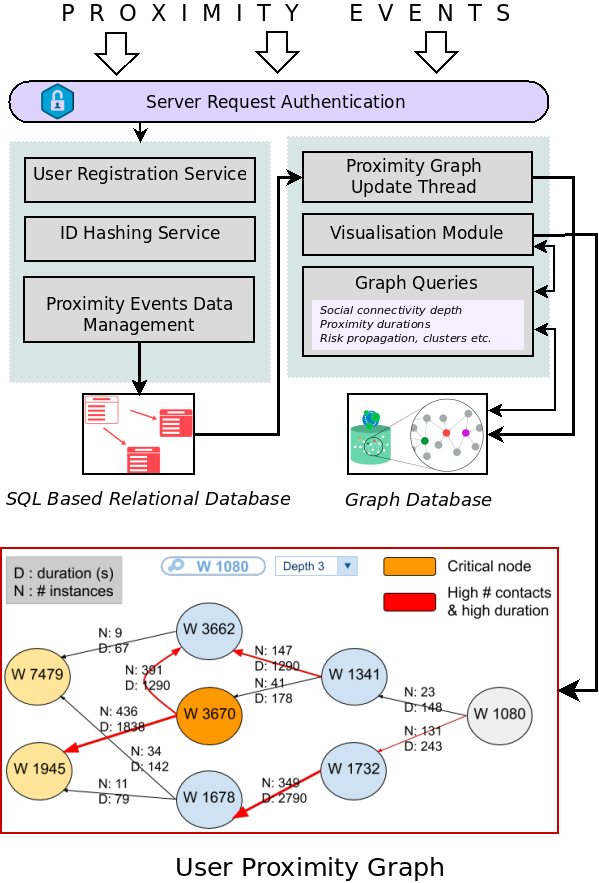}
    \caption{ProxiTrak Server Architecture.}
    \label{fig_server_arch}
\end{figure}

Figures \ref{fig_app_arch} and \ref{fig_server_arch} depict ProxiTrak's app and server side architectures highlighting the key modules.

\section{Criticality Of IT Usage}
With the advent of a new normal in light of the ongoing COVID-19 pandemic, the IT operations had shifted to predominantly virtual and online formats, with minimal personal presence in the brick and mortar workplaces. As the operations are beginning to return to normalcy at physical workplaces, a need is being established of a robust enterprise-level solution which can not only trace the spread of infection with fully customizable data privacy policies enforced by the enterprise authorities, but can also guide the workers to follow norms such as social distancing on a personal level using personal PDA devices. In addition to this, a need for multi-level contact tracing, and associate level risk profiling even in the enterprise workplaces where use of PDA devices are not allowed is also being felt. This calls for real-time proximity updates from the devices along with updating the associate's physical connectivity graph and exploiting graph structures with minimal latency. In fact, the graph analytics module and its rich inferences is proving to be the first of its kind in making IT decision making process smooth and reliable in terms of employee administration. Without the usage of IT, the aforementioned problem is extremely hard if not impossible to tackle manually.

\section{Improvement In Customer Services}
The solution holds excellent business prospects and can bag some major enterprise projects across the world (already we are on the right track with some large clients). This is because, most of the industry verticals are in requirement of solutions helping them to safely return to physical work-spaces in the wake of this pandemic. The capabilities of the solution enable it to be implemented in a variety of environments, and leverage existing BLE-tag based infrastructure or employee access management system as per need. With a good IP (Intellectual Property) portfolio backing the solution and showing ingenuity of the team, we expect this solution to yield good economic outputs. 

For the stakeholders, we expect to provide an ability of a safe return to workplaces, which will boost their operational output, with increase in employee productivity with access to workplace facilities, hardware etc., and will help normalize their revenues. The business processes can safely return back to physical work-spaces with our system in place to guide the workplaces towards an automatic enforcement of the distancing norms. The seating arrangements can be also planned using the social distancing feature, resulting in maximizing the workplace utilization. The Administration and Human Resource departments are benefiting from this solution in terms of real time monitoring, less manual intervention and automated report generation. For the end users, the solution is ubiquitous in the sense that it is a part of usual mobile device. For Health Department, this acts as an added layer of coverage, to help back track chains and connections (time interval based spatial rekations) of infected persons with BLE-enabled indoor locations (which GPS rarely covers accurately).

\begin{figure}
    \centering
    \includegraphics[width=0.85\columnwidth]{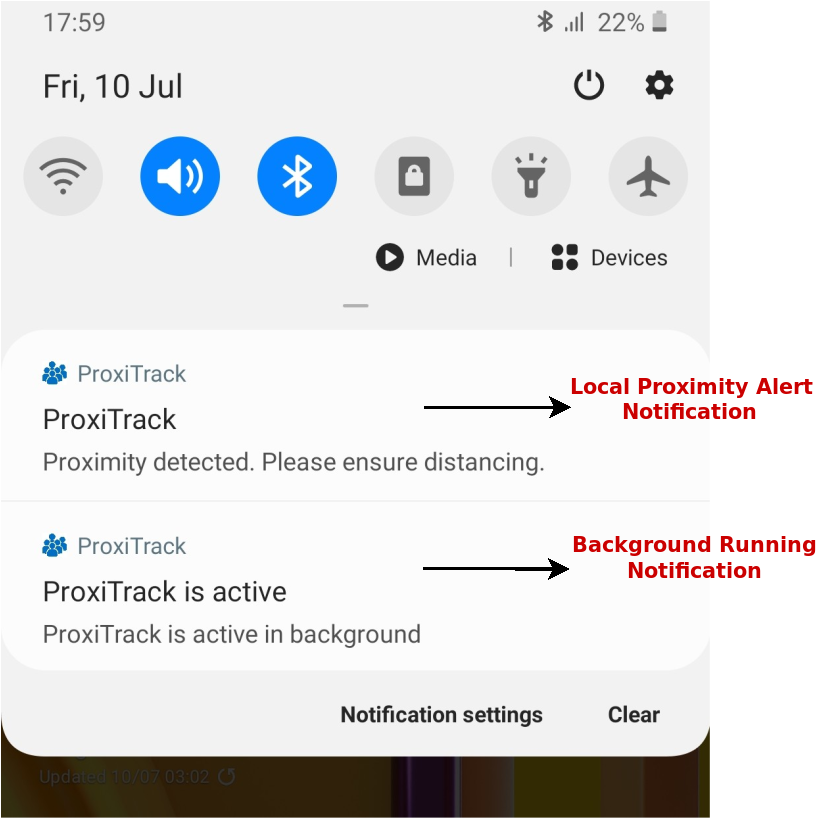}
    \caption{Screenshot of ProxiTrak app notifications showing a proximity alert, and a persistent notification showing the background status of the app.}
    \label{fig_app_screen}
\end{figure}

\section{Innovation}
The innovation was created from scratch and scaled in record time. To the best of our knowledge this is the first enterprise work for contact tracing and social distancing enablement in light of COVID-19. The problem domain addressed by our solution is unique in its own respect:

\begin{itemize}
\item Currently, there is no solution in the domain which aims at providing interventions at the social distancing stage itself.
\item Inferring proximity of associates in the regions where no personal devices are allowed is a unique problem itself in context of contact tracing.
\item Getting an acceptable accuracy in detecting proximity using a single model for multiple environments and devices is very challenging.
\item A need for fast and rich analytics for detecting risk clusters, contagion propagation etc. is endorsed by all stakeholders.
\end{itemize}

\begin{figure*}[t]
    \centering
    \includegraphics[width=6.7in]{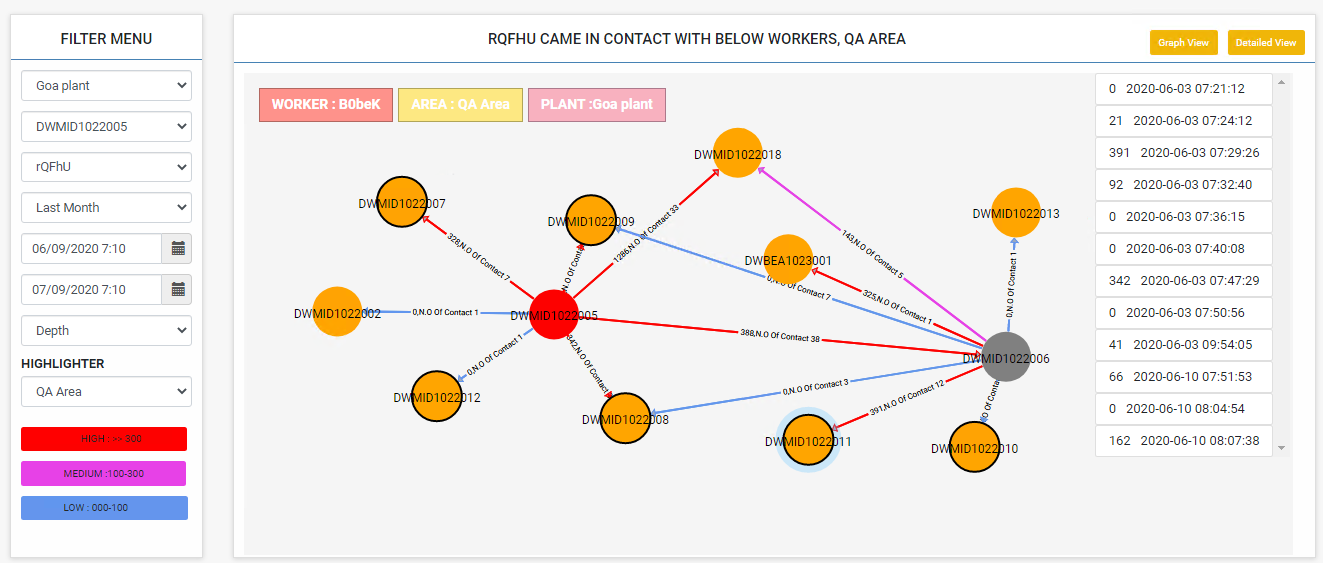}
    \caption{Graph Visualization and Contact Tracing for a real life deployment in a factory in Goa, India}
    \label{fig_graph_deploy}
\end{figure*}

Our solution is innovative on both the fronts: the app running on smart device, and the back-end server running contagion risk analysis, with a filed patent \cite{chandel2020patent} and multiple publications \cite{chandel2020}\cite{banerjee2020}:

\begin{itemize}
\item App uses a novel set of features in a machine learning model providing high accuracy in proximity detection in multitude of environments and variety of devices running different platforms.
It provides an audio-visual-tactile feedback on the device whenever any social distancing norm violation is detected.
\item Server application features multi-level contact tracing (\ref{fig_graph_deploy}), risk propagation and cluster detection powered with low-latency graph analytics. Server has been designed meticulously for scale and speed. The proximity graph is designed as a multi-graph with node and edge properties.
\end{itemize} 

\section{Return On Investment}
The project was conceived in record time with 2 resources and a mentor ($\sim2000$ person hours). This can be equated as a sort of start up venture within a MNC. The project started around mid-March 2020 and was officially launched in November 2020 as a part of digital workplace resilience solution, which is finding increased adoption among several of our B2B customers in enterprise and manufacturing domain. The following points are enlisted in terms of Return on Investment:
\begin{itemize}
	\item ProxiTrak allows our customers to reopen their business with confidence ensuring peace of mind within the enterprise setting. The post-COVID workforce management is a multi-million dollar B2B market which we are now able to tap into.
	\item Providers like airlines, banking and retail sectors can further ensure customer compliance and safety using ProxiTrak app features, which expands into the B2C market using our customer's customer base.
	\item ProxiTrak further enables a larger workforce resilience solution from TCS which can capture other logistic parameters like building management, logistic requests etc. creating a comprehensive platform for enterprises to adopt, thereby broadening the revenue horizons significantly to tap into roughly a billion dollar marker potential
	\item Proxitrak is already helping around 10,000 TCS professionals to work from office, which amounts to about 20 million USD revenue per month.
\end{itemize}

\section{Quality of Management}
ProxiTrak has been an offering from TCS Research and Innovation which got adopted as a part of business offering. In its initial days it was managed by the research group, following a very open-minded and versatile methodology for the larger good of the society in the challenging times of COVID-19. The research team had prior exposure to modeling human interactions and understanding human location context using Bluetooth on smart devices. Hence, we initiated this research on contact tracing app. However, soon the team discovered that there are multiple similar efforts going around in the world. However, the leadership was quick to identify the gap in providing real-time social distancing alerts on user devices even when they are disconnected from the Internet and multi-level contact tracing based on ambience (indoor, outdoor, air-conditioned, crowded area etc.). Hence, the team was agile to stear the research in mentioned direction and bring ProxiTrak to its current robust form.

\section{Impact on Society}
The pandemic has had a major impact on the social lives of the associates working physically away from their peers. The solution will boost the confidence of the enterprise employees assuring safety of coming back to the physical workplaces, and be able to interact face to face with their peers from a safe distance. It also enables the enterprises to accurately identify the contact traced associates thereby minimizing the associate's anxiety. A safer workplace plays a key role in enhancing the at-work productivity with an assurance of monitoring system running automatically and continuously to keep any norms' violations in check. The major social impacts are tabulated as follows:
\begin{itemize}
\item The graph analytics innovations (contagion risk measurement and propagation, risky cluster identification, contact hierarchies) can be reused easily in existing government based contact tracing systems like Aarogya Setu. This will greatly make the system more useful for administration to help in the social cause of fighting COVID-19.
\item The solution as a whole is meant to directly ensure safety and risk mitigation of humans working in offices, factories, and enterprise work spaces.
\item The solution minimizes anxiety associated with the risk of infection by digitizing the monitoring and inferences.
\item People often forget to keep social distancing due to concentration on something else or distraction; however real time alerts will help ensure safe distance is kept.
\end{itemize}
\section{Project Highlights}
ProxiTrak solution is designed to ensure safety of the associates by:
\begin{itemize} 
\item \textbf{Lowering contagion risk, Enabling safer workplaces} by generating accurate interventions (notifications) at social distancing stage itself using Machine Learning enabled proximity detections.
\item \textbf{Intelligent risk analysis} by enabling risk profiling and propagation of the contagion in a social connectivity graph, backed by low-latency database queries with real-time updates to the graph and notifications to the at-risk groups.
\item \textbf{Preservation of user \& data privacy} by using time varying unique tokens over BLE, hashing the user identities on the device, and geofencing with device-only processing of geolocation preserving both battery and location privacy. GDPR (The General Data Protection Regulation) was kept in the design from day 1.
\end{itemize}

This can be used by enterprises in any industry aiming to bring back the essential operations to physical workspaces in a safe manner.

\section{URL}
The official cloud and app deployments for multinational clients and internal Tata group can not be shared due to privacy issues and Intellectual Property held by us. Project website:

{\color{blue} \url{https://snehasisb.github.io/proxitrak/}}

\section{Technology Specifications}

The following table, lists down the technology specifications used in details.

\begin{itemize}
	\item Server - Apache Tomcat v.9
	\item Relational Database - Postgres v.12
	\item Spatial Database - PostGIS v.3 on top of Postgres
	\item Graph Database - OrientDB v.3
	\item Front-end - HTML v.5, CSS v.3, JS (jQuery v.3), vis.js
	\item Back-end - Java v.8, WAR, J2EE 8 (Servlet)
	\item Machine Learning Model - Weka (Java) - code written from scratch to enable Python library functionality in Java; transition to tensorflow framework in progress for seamless integration with both android and iOS.
	\item Server OS: Ubuntu 18.04 LTS (in Amazon AWS, MS Azure or local)
	\item Front-end Browsers: HTML5 compliant browsers like Firefox, Edge, Chrome, Opera, etc.
	\item Mobile/Tablet App: Android 5+, iOS 8+.
	\item Smartphone makes tested: iPhone, Samsung, Xiaomi, Oppo, Google Pixel, Nokia.
	
The final deployed code written for App, Client front-end and server back-end runs into 10,000+ LOC. With customization at deployment end, this code runs more as per complexity of inter-twining of existing enterprise services with this solution.
	
\end{itemize}

\section{Future Plans}
The project in its short span (as a \textbf{Make in India} product) has achieved a grand success of full scale deployment in TCS as an official app available to all 4+ lac associates as well multiple client deployments across continents and verticals (such as airline, factory, office spaces). Future plans include:
\begin{itemize} 
\item Open sourcing a version for deployment by government and non-profit administrative entities, following regulations and guidelines.
\item Pilot studies on multi-modal sensing by including other hardware sensors like camera, RFID, beacons, etc. and software. sensors like Twitter feeds.
\item Support for older versions of smartphone OS (current supports iOS 8 and Android 5 on-wards).
\item Using stream reasoning ~\cite{mukherjee2018system} on proximity events.
\item Advanced network analysis on the proximity data.
\end{itemize} 

\vspace{0.1in}
\textit{For Results and Figures - refer to cited papers below.}

\bibliographystyle{IEEEtran}
\bibliography{yitpa}

\begin{thebibliography}{1}
\providecommand{\url}[1]{#1}
\csname url@samestyle\endcsname
\providecommand{\newblock}{\relax}
\providecommand{\bibinfo}[2]{#2}
\providecommand{\BIBentrySTDinterwordspacing}{\spaceskip=0pt\relax}
\providecommand{\BIBentryALTinterwordstretchfactor}{4}
\providecommand{\BIBentryALTinterwordspacing}{\spaceskip=\fontdimen2\font plus
\BIBentryALTinterwordstretchfactor\fontdimen3\font minus
  \fontdimen4\font\relax}
\providecommand{\BIBforeignlanguage}[2]{{%
\expandafter\ifx\csname l@#1\endcsname\relax
\typeout{** WARNING: IEEEtran.bst: No hyphenation pattern has been}%
\typeout{** loaded for the language `#1'. Using the pattern for}%
\typeout{** the default language instead.}%
\else
\language=\csname l@#1\endcsname
\fi
#2}}
\providecommand{\BIBdecl}{\relax}
\BIBdecl

\bibitem{chandel2020}
V.~Chandel, S.~Banerjee, and A.~Ghose, ``Proxitrak: A robust solution to
  enforce real-time social distancing \& contact tracing in enterprise
  scenario,'' ser. UbiComp-ISWC '20, 2020, p. 503–511.

\bibitem{banerjee2020}
S.~Banerjee, V.~Chandel, and A.~Ghose, ``Graph analytics on proximity data to
  fight contagion,'' in \emph{CIKM KDAH 2020, CEUR Proc.}, vol. 2699.

\bibitem{chandel2016}
V.~{Chandel}, N.~{Ahmed}, S.~{Arora}, and A.~{Ghose}, ``Inloc: An end-to-end
  robust indoor localization and routing solution using mobile phones and ble
  beacons,'' in \emph{IPIN}, 2016, pp. 1--8.

\bibitem{ghose2013}
A.~Ghose, C.~Bhaumik, and T.~Chakravarty, ``Blueeye: A system for proximity
  detection using bluetooth on mobile phones,'' in \emph{ACM UbiComp Adjunct
  Publication}, 2013, p. 1135–1142.

\bibitem{chandel2020patent}
V.~Chandel, S.~Banerjee, and A.~Ghose, ``Real-time monitoring of proximity
  between a plurality of computing devices,'' 2020, india Patent Application
  No. 202021039568. 2020 September 13.

\bibitem{banerjee2020patent}
S.~Banerjee, V.~Chandel, and A.~Ghose, ``System and method for graph analytics
  on human proximity data to fight contagion,'' 2020, patent.

\bibitem{ghose2015patent}
A.~Ghose, A.~Jha, T.~Chakravarty, and C.~Bhaumik, ``Method and system for
  accurate straight line distance estimation between two communication
  devices,'' 2015, uS patent 9,154,904. 2015 Oct 6.

\bibitem{ghose2016patent}
A.~Ghose, S.~Arora, S.~Johari, V.~Chandel, and N.~Ahmed, ``Systems and methods
  for distance estimation and localization of users using bluetooth low energy
  beacons,'' 2016, india patent Appln no. 201621017314. 2016.

\bibitem{mukherjee2018system}
D.~Mukherjee, P.~Misra, and S.~Banerjee, ``System and a method for reasoning
  and running continuous queries over data streams,'' Jun.~5 2018, uS Patent
  9,990,403.

\end{thebibliography}

\end{document}